\documentclass[a4paper,11pt]{article}
\pdfoutput=1

\usepackage{graphicx}
\usepackage{dcolumn}
\usepackage{bm}
\usepackage{enumerate}
\usepackage{verbatim}
\usepackage{float}
\usepackage{here}
\usepackage{slashed}
\usepackage{color}
\usepackage{amsmath}
\usepackage{amssymb}
\usepackage{amsthm}
\usepackage{soul}

\usepackage{jcappub}

\usepackage[T1]{fontenc}
\usepackage{natbib}
\usepackage{amsthm}
\usepackage{slashed}
\usepackage{soul}

\title{Elucidation of `Cosmic Coincidence'}

\author[1]{Meir Shimon}
\affiliation[1]{School of Physics and Astronomy, Tel Aviv University, Tel Aviv 69978, Israel}
\emailAdd{meirs@tauex.tau.ac.il}

\date{\today}

\abstract{In the standard cosmological model the dark energy (DE) and nonrelativistic (NR) matter densities are observationally determined to be comparable at the present time, in spite of their greatly different evolution histories. This `cosmic coincidence' enigma -- also referred to as the `why now? problem' -- relies, by its very definition, on the implicit prior expectation for our `typicality' in the cosmic (expanding) spacetime volume. Otherwise, this conundrum does not exist in the first place. It is shown here that this apparent coincidence could be explained as a {\it non-anthropic} observational selection effect: for us to be typical observers 
in the comoving (static) spacetime volume, 
the cosmic energy budget must contain a non-vanishing DE 
component. In addition, it is shown that irrespective of 
the cosmological initial conditions and assuming 
no `new physics', the Universe is most {\it likely} to be 
observed at a time when the conformal Hubble 
radius, $\mathcal{H}^{-1}$, attains a maximum. 
The latter takes place 
at the epoch when $\rho_{DE}$ and $\rho_{m}$, 
the energy densities of DE and NR matter, respectively, 
are comparable. 
Specifically, our presumed `typicality' along the conformal 
timeline, coupled to a few other plausible assumptions, 
implies that $R\equiv\rho_{DE}/\rho_{m}$ 
is `sampled' from a Beta Prime probability distribution function. 
{\it A priori} 68\% (95\%) confidence range for 
the ratio is $0.20<R<3.46$ ($0.033<R<17.20$), with 
an expectation value of $\bar{R}=3.5$. 
These are in agreement with the observationally 
inferred value, $R_{obs}=2.23$.}

\begin{document}
\maketitle
\flushbottom

\section{Introduction}
\label{sec:1}
The concordance $\Lambda$CDM model provides 
a phenomenologically impressive description 
of the Universe from super-horizon scales 
down to supercluster scales with only a 
few free model parameters. These 
include the energy density of ordinary 
baryonic matter, cold dark matter (CDM) 
and dark energy (DE) [the latter 
is consistent with a cosmological constant 
$\Lambda$] along with the free parameters of the primordial matter 
power spectrum and optical depth towards the 
reionization era. 
However, the standard $\Lambda$CDM model is still marred by several 
long-standing puzzles.

Whereas the consensus within the community, 
given currently available observational data, 
is that the existence of some form of vacuum-like 
DE seems to be unavoidable, little is known about what 
it might actually be, e.g.
\cite{3,15,22,32,33,34,39}.
Fundamentally, it is widely believed that the vacuum-like 
energy density is induced by some unknown 
slow-rolling scalar field which is not part of 
the standard model (SM) of particle physics, or effectively by a 
cosmological constant. Regardless (or not) 
of its true identity, two major problems are 
raised by the apparent need for DE.  

First, from very general considerations, 
the zero-mode fluctuations are expected 
to induce a vacuum energy density at the 
$O(1)M_{p}^{4}$ level, e.g. \cite{3}, where $M_{p}$ 
is the Planck mass, and the lifetime of a 
Universe dominated by this vacuum energy is 
the Planck time. 
It has long been hoped that due to some 
(yet-unknown) symmetry principle 
the vacuum energy density identically 
vanishes. This all changed in 1998 when it 
was discovered that cosmic expansion 
started accelerating at redshift $\lesssim 1$ \cite{1998,1999}.
Nearly $69\%$ of the 
cosmic energy budget at present is 
accounted for by a vacuum-like DE, 
consistent with a cosmological constant whose 
energy density is $\sim 122$ orders of magnitude 
smaller than the Planck density, in 
sharp disagreement with theoretical expectations. 

A second problem related to 
the cosmological constant -- 
the Cosmic Coincidence problem (CCP) -- is the 
observational fact that nonrelativistic (NR) 
matter and vacuum energy densities are of 
the same order of magnitude at present, 
whereas their evolution histories 
are very different \cite{114}. Specifically 
the DE-to-NR energy ratio scales 
$\propto(1+z)^{-3}$, and for their 
ratio to be unity at present it 
must have been fine-tuned at, 
e.g. $\gtrsim 80$ decimal places 
at the GUT epoch. For comparison, 
the spatial curvature of the Universe 
must have been fine-tuned at the GUT 
era at 27 decimal places for the 
Universe to be nearly flat at present. 
This latter fine-tuning, known as 
the `flatness problem', was a major 
thrust for introducing an early 
inflationary era to the cosmological model.
To illustrate the absurdity 
of this result, imagine that NR and DE are 
not equal at present, but rather that the energy 
density of DE at present is 20 orders of magnitude 
smaller than that of NR (regardless 
of the fact that this tiny value is unlikely to 
be detected). Even then, 
the required fine-tuning is at the stupendously 
large, $\gtrsim 60$ orders of magnitude, level. In other 
words, the CCP is 
only one particular aspect of an even more 
perplexing puzzle -- `why does DE exist at all?'

An exactly identical reasoning could be applied to the flatness 
problem, with the inevitable conclusion that spatial curvature 
should exactly vanish, unless unacceptable level of fine-tuning of 
the initial conditions is allowed. However, cosmic inflation allow the flattening of space and the subsequent materialization of energy 
such that space within the observable Universe is very nearly flat.

The CCP, sometimes referred to as the `why now?' 
problem, can alternatively be posed as follows: 
Whereas the Hubble radius is 
generally time-dependent, 
the length scale, $l_{\Lambda}=1/\sqrt{\Lambda}$, 
associated with the cosmological constant, $\Lambda$,  
is also a constant. 
The problem is then to explain how is it that specifically at the time of 
observation, i.e. now, these two scales are comparable. 

All these different formulations of the CCP are at best qualitative, 
citing the number of decimal places of the 
required precision in setting the initial conditions 
rather than referring to a quantitative statistical estimate of tension with theory. The reason for that 
is that theory does not tell us what values should 
the various cosmological parameters take. Some 
clues are provided by toy quantum cosmology models 
or the multiverse as to the likely range of values 
of the various parameters but these heavily depend 
on the assumed measure, e.g. \cite{49, 102, 103, 104, 111}.
Consequently, we find ourselves in the uncomfortable 
situation that on the one hand the observed cosmic coincidence 
seems to be puzzling and {\it a priori} unlikely, but on the 
other hand this unlikeliness is not well-defined or quantified 
in a statistically precise fashion.

Proposed resolutions of the CCP range from anthropic 
considerations, e.g., 
\cite{2,3,4,6,9,12,14,19,24,35,37,46,48,52,55,58,105, 106} through tracking models \cite{7,8,10,13,75}
to interacting DE-DM models 
\cite{20,26,28,38,67,72}.
Other explanations, e.g. 
\cite{16,17,18,21,23,27,29,30,31,
41,42,47,50,56,62,69,70,73,74}, abound. 
While widely believed to be a problem, there is also 
the opposing view maintaining that there is no 
problem to begin with, e.g. \cite{5,60}.

In this work an explanation is proposed of 
the currently comparable values of 
DE and NR energy densities as a non-anthropic 
{\it observational selection effect}: 
assuming that we are `randomly selected' observers 
in the comoving (static) -- rather than in 
cosmic (expanding) -- spacetime volume it is shown that we are 
most likely to observe the Universe when $R=O(1)$, 
where $R$ is the ratio $R\equiv\rho_{DE}/\rho_{m}$, and 
$\rho_{DE}$ and $\rho_{m}$ are the energy densities of 
DE and NR matter, respectively.
This takes place at the minimum total energy density, or 
equivalently, at the maximum Hubble scale, or minimum 
(conformal) expansion rate of the Universe, when the 
Universe spends a relatively large fraction of its 
conformal lifespan. It should be stressed already from 
the outset that whereas the assumption that we are 
randomly sampled in the comoving spacetime volume 
could be arguably considered to be {\it ad hoc}, a similar assumption 
about our typicality in the cosmic frame, 
(over a sufficiently long, but finite, period of time) 
lies at the foundations and even existence of the CCP, 
as well as other problems, e.g. the `Boltzmann Brain' 
problem of the standard $\Lambda$CDM model, e.g. \cite{121, 115}. 
In the absence of any guiding principles from theory, the 
ultimate test for any such prior expectation is the naturality or unnaturality 
of the observationally inferred cosmological parameters given any particular prior assumption. In this work we entertain the plausible possibility that the CCP (among other problems of the $\Lambda$CDM model) does not exist if we are randomly `drawn' in the comoving -- rather than the cosmic -- frame.

The structure of the paper is as follows. 
In section \ref{sec:2} we advocate the choice of a 
uniform prior in the comoving spacetime volume instead of 
the default analogous prior in the cosmic spacetime volume. 
This is not essential for our likelihood considerations in 
section \ref{sec:3} (the latter by no means relies on the arguments that are outlined in section \ref{sec:2}) where our analysis 
is carried out in the `dual' redshift space. 
In section \ref{sec:4} additional implications are considered beyond the narrow interest of the CCP. We conclude with a summary in section \ref{sec:5}.   

\section{Typicality in the Comoving vs. Cosmic Frame}
\label{sec:2}
By their very essence, `why now?' problems in cosmology 
owe their existence to the implicit assumption that 
there is {\it a priori} no preferred time scale in our 
cosmological models to determine that, e.g. the 
energy densities associated with DE and NR matter 
should be comparable specifically at the time of 
observation. In other words, 
if anthropic reasoning is excluded, then it is 
assumed that we are essentially `drawn' from a uniform 
prior distribution on the time variable. 
The essence of the `why now?' puzzle is 
the following; under that assumption, how is it that we are 
`randomly selected' to observe at just this {\it special} 
epoch when DE and NR (energy) densities are comparable in 
spite of their very different evolution histories. 
This seems to require stupendously fine-tuned initial 
conditions, otherwise the comparability of DE and NR (energy) 
densities at the present time, of all times, seems 
very {\it improbable}.

While in general there is a clear sense that this needs 
to be addressed, the problem itself is not precisely 
quantified. In particular, the standard cosmological model 
does not inform us what should the model parameters empirically 
be, even not statistically. For example, the present-day 
energy densities of DE and the NR matter can {\it a priori} 
take any value and it is only 
experiment/observation that pins them down. Therefore, it is impossible 
to quantify, in a conventional statistical way, the severity of 
the problem in terms of, e.g. statistical `tension' between theory and 
observations. 
Rather, one is left with estimates, such as those described in section \ref{sec:1}, of the severity of fine-tuning at arbitrary times, e.g. 
GUT time, or the time when inflation ended, or any other time, in terms 
of number of decimal places -- we simply 
do not know when exactly the initial conditions have been set, but admittedly 
these sort of estimates provides some notion of the severity of the problem. 

In addition, in asking why `now' of all times are 
the DE and NR matter densities comparable, one necessarily 
assumes something about time. The time coordinate 
is assumed by default to be cosmic time, $t$. However, 
a uniform prior on $t$ is not equivalent to a uniform 
prior on, e.g., conformal time, $\eta$, and there is no 
{\it a priori} reason to favor the former over the latter, 
or over other priors on any time coordinate, $\tilde{t}$. 
The conformal time, $\eta$, 
will be defined below. In the absence of any additional 
evidence the question of what prior 
to use is ultimately decided in practice 
by comparison of theoretical predictions 
with observations, guided by 
the principle that the prior which is most 
naturally compatible with observations, i.e. 
with the least amount of apparent fine-tuning, 
should be favored over others. 

One purpose (among others) of the present section is to advocate 
the choice of a uniform prior on the comoving (static) rather 
than cosmic (expanding) spacetime volume (which for the time 
coordinate implies uniform prior on $\eta$ rather 
than $t$). However, as mentioned above, the choice 
can be made {\it ad hoc}, much like it is tacitly assumed conventionally 
(with no clear justification) that $t$ is drawn from a uniform prior (over a sufficiently long, but finite, time period) with no evidence to support this assumption. Quite the contrary, a uniform prior on $t$ is impossible 
(but rather only over a finite period) and it results in a few 
well-known puzzling properties of the standard 
cosmological model, primarily the CCP.

How likely is it for us to make observations of an eternal 
Universe at the first 14 Gyrs of its existence? 
This puzzle that pertains to a Universe 
that has a beginning but no end, such as the Universe we 
seem to inhabit, implies that there must be a finite 
probability for our existence within a finite time interval 
after the Universe `starts'. Our prior position along 
the cosmic timeline, i.e. along the $t$ axis, 
cannot be uniform in an eternal Universe, thereby 
making us `special', with pronounced preference to 
observe the Universe at a measure zero fraction 
of its eternal lifespan, e.g. closer to its beginning than to 
its (infinitely remote) end. This `naturality' issue does 
not exist in Universes which are both past- and future-eternal 
on the one extreme, and in Universes that end in, e.g. a Big Crunch, 
Big Rip \cite {109}, or Big Slurp (sudden decay of false vacuum state) \cite{110}, on the other extreme. 

According to Laplace's Principle of Insufficient Reason, also known as 
the Indifference Principle (that was also 
discussed by Bernoulli, Poincare, and many others), 
in the presence of maximum ignorance as to how likely is it 
for an event to take place -- all possibilities are equally likely. In Bayesian statistics, this is the simplest 
non-informative (`maximum entropy') prior probability 
-- a uniform distribution. Let us assume that the {\it true} 
prior probability of observing the Universe over an infinitesimal 
time interval, $dt$, around any given {\it cosmic} time, $t$, is $\mathcal{P}(t)dt$, where $\mathcal{P}(t)$ 
cannot be a non-vanishing constant since according to 
the concordance cosmological model the Universal expansion is eternal into the future. A new time coordinate, 
$d\tilde{t}\propto\mathcal{P}(t)dt$, can always be so defined 
that in terms of which the new probability distribution 
$\tilde{\mathcal{P}}$ is uniform, i.e. 
$\tilde{\mathcal{P}}d\tilde{t}=\mathcal{P}(t)dt$.
Let us also assume that $\tilde{t}\in[0,T]$, 
where $T$ is finite.
The latter assumption is critical, because it guarantees that observations 
carried out over a {\it finite} interval, 
$\Delta\tilde{t}$, 
however small it is, have non-vanishing finite probability 
to actually take place. In this new frame half of the typical 
observers, i.e. those who were `randomly sampled' from a flat prior in the range $[0,T]$ will observe the Universe 
over the middle time 
interval of the entire duration $T$, i.e. $\tilde{t}\in[T/2,3T/4]$, 
$68\%$ percent of them will observe at $\tilde{t}\in[0.16T,0.84T]$, 
$95\%$ percent of them will observe over the range 
$\tilde{t}\in[0.025T,0.975T]$, etc. We stress once again that the objective of the (admittedly somewhat simplistic) analysis carried out in this section is only to motivate our choice $\tilde{t}=\eta$, and the more 
detailed and rigorous analysis is relegated to section \ref{sec:3}.

In the following, we apply the Principle 
of Indifference to observations made in a Universe like ours, 
which is described by the standard cosmological model. 
The most general homogeneous and isotropic spacetime 
is described by the Friedmann-Robertson-Walker (FRW) metric
\begin{eqnarray}\label{eq:2.1}
ds^{2}=-dt^{2}+a^{2}(t)
\left[\frac{dr^{2}}{1-Kr^{2}}+r^{2}(d\theta^{2}
+\sin^{2}\theta d\varphi^{2})\right]\nonumber\\
=a^{2}(\eta)\left[-d\eta^{2}+\frac{dr^{2}}{1-Kr^{2}}+r^{2}(d\theta^{2}
+\sin^{2}\theta d\varphi^{2})\right], 
\end{eqnarray}
where $t$ and $\eta\equiv\int\frac{dt}{a}$ are the cosmic and conformal 
time coordinates, respectively, $a(t)$ is the scale factor, and $r$, 
$\theta$ and $\varphi$ are the standard spherical coordinates. 
The spatial curvature parameter is $K$.
In the `t-frame' [first of Eqs. \eqref{eq:2.1}] the lapse 
function is normalized to unity \cite{1}.
It is customary to set $a=1$ at 
the present time. In this convention $a=(1+z)^{-1}$, 
where $z$ is the cosmological redshift. The second 
of Eqs. \eqref{eq:2.1} describes 
the same spacetime in a more `symmetric' form where $a(\eta)$ 
plays the role of an overall conformal factor. In the following 
we refer to this frame as the `$\eta$-frame'. 
The frame in which $a(\eta)$ is scaled out is known as 
the `comoving frame', essentially a static spacetime 
described by the metric 
$g_{\mu\nu}={\rm diag}(-1,\frac{1}{1-Kr^{2}}, r^{2}, r^{2}\sin^{2}\theta)$  where masses rescale as $m_{0}\rightarrow m_{0}a(\eta)$. According to 
this alternative picture the entire cosmic evolution is 
manifested by growing masses on a static background spacetime.
These two alternative pictures are observationally indistinguishable, 
but as we see below (and elsewhere \cite{122}) a uniform prior on events in the comoving spacetime volume is favorable over a uniform prior on events in the cosmic spacetime volume from the perspective of naturalness and typicality. 

The Copernican Principle, referred to as the Cosmological Principle 
when applied to the cosmological model, embodies the idea that 
all observers at a given cosmic (or conformal for that matter) 
time are `typical'. 
However, such a uniform prior on spatial location 
only results in non-vanishing probabilities if 
the causally connected volume, i.e. 
the observable Universe, is finite. 
This is possible if, e.g. space has a closed 
geometry or, alternatively, 
if there is a cosmological horizon that will never 
be crossed, even in the very remote future.
The latter forms when cosmic expansion enters 
an accelerated phase.
As mentioned above, naively applying the 
Copernican Principle to (cosmic) 
time itself is problematic in our Universe since 
the $t$ coordinate is future-infinite. However, 
if cosmic horizons that result 
from accelerated expansion do exist then $\eta$ is finite. 
We then conclude that the Copernican Principle 
can be applied to spacetime, 
not only to space, if cosmic expansion enters an 
acceleration phase. 
In that case we can in principle think 
of uniform priors on each 
one of the spacetime coordinates, 
with events parameterized by 
$(\eta, x, y, z)$, where $\eta$, $x$, $y$ and $z$ 
are independent coordinates in the comoving 
frame. This four-dimensional uniform distribution -- 
which is the one advocated in the present work -- is 
independent of space expansion because it applies 
to the comoving coordinates; space expansion is 
replaced in this alternative description by 
monotonically growing masses.

The temporal evolution of $a(t)$ 
is governed by the Friedmann equation 
\begin{eqnarray}\label{eq:2.2}
H^{2}=\frac{8\pi G\rho}{3}, 
\end{eqnarray}
where $G$ is the Universal gravitational constant, 
$H\equiv d\xi/dt$ is the Hubble function where $\xi\equiv\ln a$ 
has been defined, and $\rho=\sum_{i}\rho_{i}$ is a sum over all 
contributions to the total energy density, including DE. 
The conformal Hubble function is $\mathcal{H}\equiv d\xi/d\eta=aH$. 
In the case that these species do not mutually interact 
and each one individually satisfies the continuity equation, 
then Eq. \eqref{eq:2.2} can be written as
\begin{eqnarray}\label{eq:2.3}
H^{2}(t)=H_{0}^{2}\sum_{i}\Omega_{i}e^{-3(1+w_{i})\xi}, 
\end{eqnarray}
where $H_{0}$ is the Hubble constant, 
$\Omega_{i}$ is the energy density of the i'th species 
in critical density units, $\rho_{c}\equiv 
3H_{0}^{2}/(8\pi G)$, $w_{i}$ are the respective 
equations of state (EOS), and the various $\Omega_{i}$'s 
are subject to the constraint $\sum_{i}\Omega_{i}=1$ 
(the effective energy density of curvature, $\Omega_{k}=-K/H_{0}^{2}$, 
included). Clearly, an observation cannot be made by sentient 
beings (to the best of our current understanding) during the RD era 
or cosmic inflation, and so our primary focus in this work is 
on the post-RD era (however, we do not have to rely on this anthropic 
consideration in the case of the concordance model -- this point is referred to below). Spatial curvature is also ignored for 
the most part of this work, and so $\Omega_{DE}=1-\Omega_{m}$, 
where $\Omega_{m}$ denotes the energy density of NR matter in 
critical density units. 

Integration of the Friedmann equation, 
assuming only non-relativistic (NR) matter and and DE (the latter is characterized by an EOS $w$) 
results in the conformal time lapse (in Hubble time units) from the big bang to any desired redshift $z$
\begin{eqnarray}\label{eq:2.4}
H_{0}\eta_{z}\equiv H_{0}\Delta \eta(z)=\int_{\infty}^{z}\frac{dz'}{\sqrt{\Omega_{m}(1+z')^{3}+(1-\Omega_{m})(1+z')^{3(1+w)}}}. 
\end{eqnarray}
Applying \eqref{eq:2.4} to the case $z=-1$ 
in a Universe purely made of NR matter results in
divergent cosmic lifetime as measured in terms of both $t$ and $\eta$.
This divergence comes from contributions near $z\rightarrow -1$.
However, adding a `DE' component (with $w<-1/3$), 
that drives an accelerated expansion when the NR 
contribution becomes subdominant to DE still 
results in divergent $\Delta t$ over the lifetime of the Universe, but 
a finite
\begin{eqnarray}\label{eq:2.5}
H_{0}\eta_{-1}/2=\frac{\Gamma(1-\frac{1}{6w})\Gamma[\frac{1}{6}(3+\frac{1}{w})]R^{\frac{1}{6w}}}{\sqrt{\pi\Omega_{m}}},  
\end{eqnarray}
where $R\equiv\Omega_{DE}/\Omega_{m}$, and $\Gamma(x)$ 
is the Gamma-function. 
The (dimensionless) conformal lifetime $H_{0}\eta_{-1}$ 
is finite due to the existence of an asymptotic future 
{\it cosmic horizon}. 
The limiting case $w=-1/3$ results in a logarithmic divergence.
In other words, any species characterized by $w<-1/3$ 
`regularizes' the (otherwise) infinite (conformal) lifetime 
of the Universe.
The present conformal cosmic age is given by
\begin{eqnarray}\label{eq:2.6}
H_{0}\eta_{0}/2=\frac{{_{2}}F_{1}(\frac{1}{2}, -\frac{1}{6 w}; 1-\frac{1}{6w}, -R)}{\sqrt{\Omega_{m}}}, 
\end{eqnarray}
where ${_{2}}F_{1}$ is the hypergeometric function.

\begin{figure}[h]
\begin{center}
\leavevmode
\includegraphics[width=0.5\textwidth]{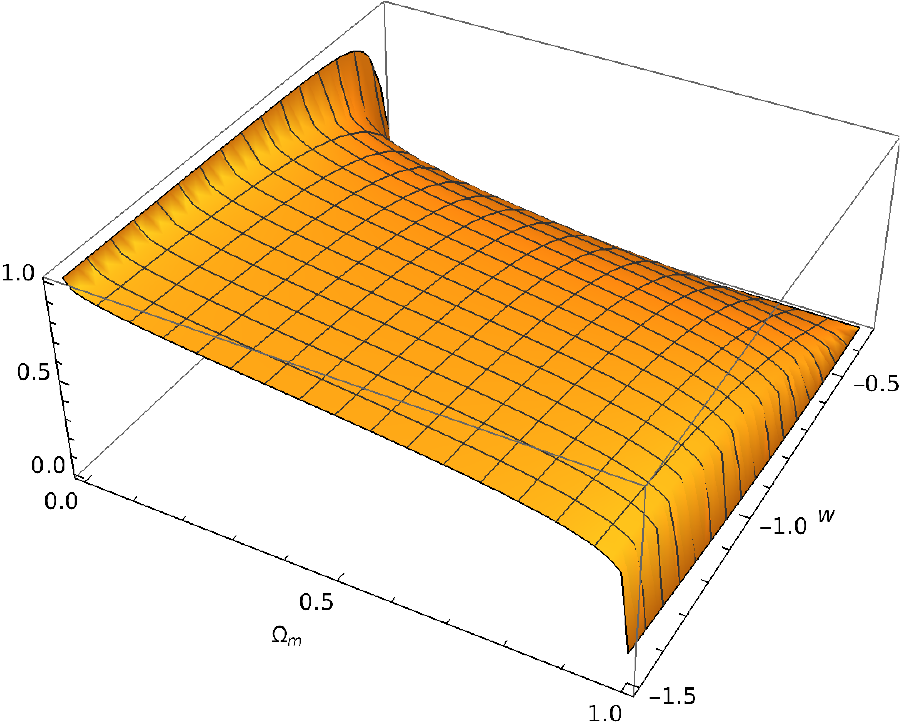}
\end{center}
\caption{\label{fig:1}
The conformal age of the Universe in conformal lifespan units 
($\eta_{0}/\eta_{-1}$) for a range of $\Omega_{m}$ and $w$ 
values: Unless $w\rightarrow -\frac{1}{3}$ or $\Omega_{m}\rightarrow 1$ (where there is basically no horizon and $\eta_{-1}$ diverges) 
the ratio $\eta_{0}/\eta_{-1}$ is a sizable fraction of unity, 
namely we are {\it a priori} bound to find that we are 
`average' observers. In the cases of very negative $w$ the 
Universe is short-lived and we are likely to find ourselves in an old Universe. A similar effect is found when $\Omega_{DE}\rightarrow 1$, 
i.e. when the Universe exponentially expands and so $\eta_{0}/\eta_{-1}\rightarrow 1$.}
\end{figure}

\begin{figure}[h]
\begin{center}
\leavevmode
\includegraphics[width=0.5\textwidth]{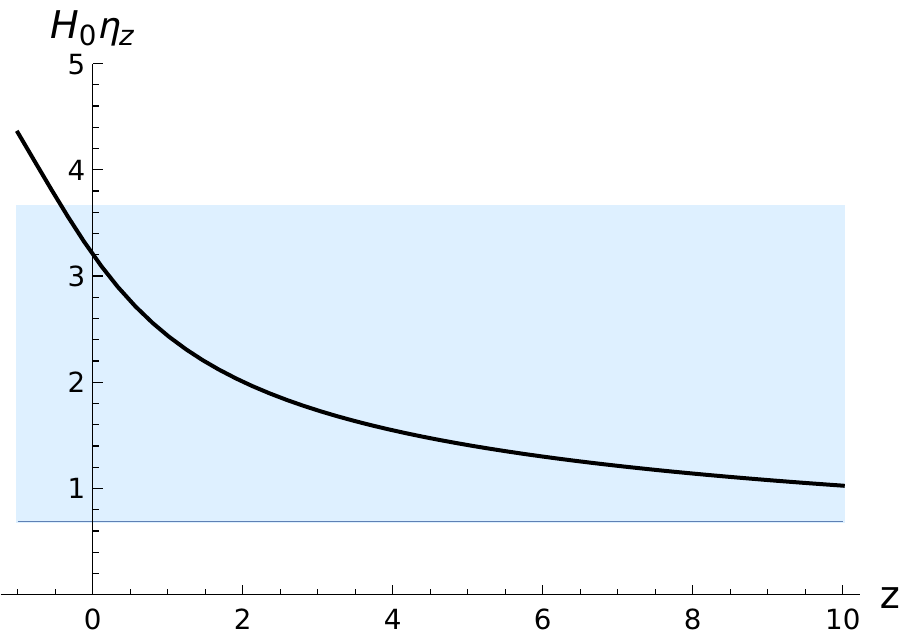}
\end{center}
\caption{\label{fig:2}
The conformal time lapse in Hubble time units, 
$H_{0}\eta_{z}$, is plotted 
for a range of redshifts, assuming the concordance $\Lambda$CDM 
model parameters. The middle 68\% range of 
[$H_{0}\eta_{\infty}$,$H_{0}\eta_{-1}$] 
is highlighted as a shaded light blue region. The present 
day $H_{0}\eta_{0}$ value falls within this `confidence' region.}
\end{figure}

The ratio $\eta_{0}/\eta_{-1}$ is generally a sizable 
fraction of unity for a wide range of $w$ and $\Omega_{m}$ values as 
can be seen from Fig.~\ref{fig:1}.
Specifically, applying Eqs. \eqref{eq:2.5} -- \eqref{eq:2.6} to 
the concordance $\Lambda$CDM model with $\Omega_{m}=0.31$ 
and $w=-1$ results in $\eta_{0}/\eta_{-1}=0.74$, 
i.e., we are well within the middle 68\% confidence 
interval in terms of $\eta$ [according to the prescription 
outlined above Eq. \eqref{eq:2.1}], and from this narrow perspective we 
can indeed be considered average observers; we observe the Universe at a point where 3/4 of the lifetime of our Universe has 
passed (as opposed to the cosmic frame where we are 
very atypical observers who just happen to observe the 
Universe 14 Gyrs after it formed, essentially at 
time zero compared to its eternal lifetime). 
According to the concordance cosmological model 
$\eta_{0}$ and $\eta_{-1}$ equal 45 and 61 (conformal) 
Gyrs, respectively, i.e. we will inevitably 
hit the `end of time' in 16 (conformal) Gyrs from now.
This is shown in Fig.~\ref{fig:2} where $H_{0}\eta_{z}$ is plotted 
for a range of redshifts. The shaded light blue region corresponds to the middle 68\% range of $H_{0}\eta_{z}$ ($-0.42<z<21.9$), and the 
median takes place at $z=1.54$. The present 
day $H_{0}\eta_{0}$ value falls within this range.
For reference, the middle 95\% range lies within the 
range ($-0.91<z<528$), essentially excluding the 
radiation-dominated (RD) era (radiation-matter 
equality takes place at $z_{eq}\approx 3400$ 
and the {\it a priori} betting odds for us to 
be randomly drawn from a uniform distribution 
in $\eta$ at the RD era are less than $7$ in 
a thousand). In other words, given that 
$\Omega_{m}=0.31$, and assuming that we are typical 
observers in the comoving frame excludes the possibility 
that we are formed at the RD era at $99\%$ confidence 
with no recourse to anthropic reasoning, biological 
evolution modeling, etc.  

The conclusion of all this is that whereas the assumption 
that we are randomly positioned in the cosmic (expanding) 
spacetime volume of a Universe that has 
a beginning but no end results in glaring typicality problems, 
{\it the alternative hypothesis that we are observers 
who are `randomly selected' from a uniform prior distribution 
in the comoving spacetime volume results in no 
such naturalness problems}, provided that at some point the 
Universal expansion (as described in the expanding frame) 
accelerates. In other words, for 
us to be typical in the comoving spacetime volume 
plausibly requires DE to exist. 
In the next section we apply a uniform prior 
in $\eta$ to the CCP formulated in redshift-space.

We conclude this section with a demonstration 
of the utility of the assumption of 
`typicality' in the comoving spacetime volume.
Specifically, and with an eye towards the discussion in 
section \ref{sec:3}, 
useful bounds on the ratio $R$ are derived (with no recourse to any 
prior knowledge) under 
the single assumption that we are randomly 
situated in the comoving spacetime volume 
of a Universe that has a finite 
conformal lifetime (i.e. it ends its evolution with an 
accelerated expansion phase). 
Specifically, we customize an approach, that was 
popularized by Gott originally in the context of the 
`Doomsday Argument' \cite{100}, to the case at hand.
This, essentially typicality (sometimes referred 
to as `mediocrity') argument, e.g. \cite{102, 100}, is not 
uncontroversial, e.g. \cite{46, 108, 112, 113}, but it is nevertheless 
useful, in the spirit of the present section, 
to derive these constraints from the assumption that 
we are typical in the comoving frame. This is especially 
warranted in the context of the cosmic `why now?' problem, 
which is at the focus of the present work; 
as mentioned above, in its very essence this problem arises 
due to the tacitly made assumption that we are typical observers 
along the cosmic timeline $t$, or in other words it is implicitly assumed that we are typical in the expanding -- cosmic -- frame, i.e. why 
are $\Omega_{m}$ and $\Omega_{DE}$ comparable at present ($t_{0}$) while 
they are much different at any other time; what makes 
the present time particularly special (given the prior expectation that 
we are typical)?

Assuming that we are random observers 
along the Universal $\eta$-axis then there is 50\% chance 
that we are in the middle 50\% interval of the entire 
Universal lifespan. The two extremes of this confidence interval 
are at 1/4 and 3/4 the way from the big bang, $\eta_{\infty}=0$, 
to the end of time, $\eta_{-1}$. If we happen to observe the Universe 
at the one extreme, then $\eta_{0}=\frac{\eta_{-1}}{4}$. 
On the other hand, if we observe the universe at the other extreme 
of this interval, then $\eta_{0}=\frac{3\eta_{-1}}{4}$. 
Inverting these relations we obtain
$\frac{4}{3}\eta_{0}<\eta_{-1}<4\eta_{0}$ at 
the 50\% confidence level.
Applying a similar reasoning to obtain the 68\% confidence 
interval results in $1.19\eta_{0}<\eta_{-1}<6.25\eta_{0}$. 
Employing Eqs. \eqref{eq:2.5} -- \eqref{eq:2.6} we obtain that $R\in[0.00186,2.55]$ and 
$R\in[1.128\times 10^{-4},10.70]$ at the 50\% and 
68\%\ C.L, respectively.
These estimates, which rely on no prior measurements or 
data, and only assume that we are typical observers along a {\it finite} 
$\eta$ axis as well as the validity of Eqs. \eqref{eq:2.5} 
-- \eqref{eq:2.6} 
(i.e. the assumption that the Friedmann equation adequately 
describes the cosmic evolution and that 
typical observers are found in a Universe largely 
dominated by either 
NR matter, DE, or both) should be compared to the 
observationally inferred value. 
According to the concordance $\Lambda$CDM 
model $\Omega_{m}=0.31$ and so $R_{obs}=2.23$. Since these probability 
estimates rely on our {\it a priori} ignorance as to our actual 
position in {\it spacetime}, rather than on any prior knowledge, the 
existence of DE could have been predicted long before the 
discovery of the accelerated expansion in 1998 \cite{1998,1999}, 
if only we were assumed to be `randomly sampled' from the 
comoving static spacetime volume. 

It is startling that applying the 
`Principle of Indifference', basically assigning uniform 
prior probability to our position in the comoving 
spacetime volume, is sufficient to conclude that a 
DE component (described by $w<-1/3$) must exist. 
The observed ratio, $R_{obs}$ is then found to be consistent with 
the $68\%$ confidence interval that is retrodicted 
based on this remarkably modest assumption. However, it has 
been claimed that the seemingly innocuous assumption of typicality grants us an (perhaps unwarranted) `enormous leverage' \cite{108, 115}.
On the other hand, and specifically following the argument 
above Eq. \eqref{eq:2.1}, the assumption of typicality is 
always correct is some specific time coordinate $\tilde{t}$. 
In the present work we advocate that this time coordinate 
is the conformal time, $\eta$.

All these considerations and results lend support 
(but are not essential) to our proposal to replace the 
implicit assumption of typicality in the cosmic frame 
with a more `realistic' typicality in the comoving frame. 
Again, it is certainly legitimate to {\it ad hoc} posit a prior 
on $\eta$ in the same fashion that it is conventionally 
and tacitly done for $t$ (over a sufficiently long period 
after the big bang). 

\section{Observational Likelihood Function}
\label{sec:3}
Proposed anthropic solutions to 
the CCP invariably involve considerations 
of the growth of structure. However, for our 
non-anthropic arguments it will be 
sufficient to work at the smooth and homogeneous 
background level. 
The analysis of section \ref{sec:2} was restricted 
to the time-domain. In the present section we carry 
out our probabilistic estimates in the `dual', 
redshift-domain, taking advantage over the entire available 
redshift range that provides a better leverage and constraints 
on the energy density parameters. The two are related by gravitation, 
specifically via the Friedmann equation in our case. 

Before turning to the construction of the likelihood function in redshift 
space it is perhaps constructive to examine a well-known 
mechanical system where a similar observational selection 
effect is most easily grasped -- that of a one-dimensional 
undamped harmonic oscillator. 
If the periodic motion of an oscillator is randomly 
sampled in time, it is more likely to be found near the extreme 
ends of the range of its motion (where it momentarily comes to 
an halt just before turnaround) than near the equilibrium point where the motion is the fastest. A straightforward calculation 
shows that it only spends 1/3 of each oscillation period 
(i.e. of the number of random samplings) in the middle half 
of its range of motion  between the extreme points. 
The rest 2/3 is spent in the far half of its trajectory, 
closer to the end points. In this classic observational selection effect 
example the oscillator instantaneous velocity plays an analogous role 
to the one played by $\mathcal{H}$, the conformal Hubble expansion rate, 
in the cosmological context as we see below.

Back to cosmology, following the discussion in 
section \ref{sec:2}, and adopting a uniform 
prior in the comoving spacetime volume rather 
than expanding volume, our `position' 
along the timeline is `drawn' from the prior
\begin{eqnarray}\label{eq:3.1}
\mathcal{P}(\eta)d\eta=Cd\eta, 
\end{eqnarray}
where $C^{-1}\equiv\eta_{-1}$ is the lifetime 
of the Universe as defined by Eq. \eqref{eq:2.5}. 
In terms of cosmic time, $t$, this prior corresponds to 
a probability distribution $\mathcal{P}(t)\propto a^{-1}$. In an 
expanding Universe that grows with no limit this 
distribution favors `small' scale factors, $a$, i.e. 
earlier cosmic times in an eternal Universe. 
In other words, although our Universe is likely to spend 
essentially an infinite time in an exponentially, DE-dominated, 
phase, we still happen to observe it at just the 
transition phase between matter and DE-domination. This is because the odds are systematically tilted towards smaller scale factors, i.e. earlier times, due to the probability 
distribution $\mathcal{P}(t)\propto a^{-1}$, as is proposed by Eq. \eqref{eq:3.1}, that effectively 
imposes an exponential cutoff once the accelerated 
expansion phase begins; we observe the Universe at the latest 
(practically) possible time in an infinite $\Lambda$CDM Universe. This provides a heuristic argument, under the assumption of flat prior in $\eta$, for the fact that our observation is made during the transition epoch between NR matter and DE; while it is true 
that the Universe spends an infinite time period 
in the DE-dominated era the probability of making an observation during this period is exponentially suppressed. Since the transition takes place a finite 
(cosmic) time after then big bang then it stands to 
reason that observations are most likely to take place 
a finite (cosmic) time after the big bang in spite of the fact that 
the Universe is eternal.

Consider the concordance $\Lambda$CDM model in which 
the cosmic energy budget is dominated 
by a vacuum-like energy density in the 
far future. The conclusion of the following 
discussion is robust to reasonably 
changing the EOS describing the asymptotic future 
to any constant $w<-1/3$ or even slowly evolving EOS. 
We will see below that a 
temporally random observer (in $\eta$) is most likely 
to make observations when $R=O(1)$. We stress once again that 
matter perturbations over the smooth background are not 
required to satisfy any particular conditions (or even 
to exist) for the following considerations to apply. 

Observing the Universe over a 
finite (conformal) time interval $\Delta\eta$ has a non-vanishing 
probability {\it only if} $\eta$ is finite, 
i.e. the probability density function is normalizable. 
As mentioned above, our very existence a 
finite (cosmic) time after the big bang in an eternal Universe 
is already very puzzling if we were to be drawn 
from a {\it uniform} prior on $t$ unless 
anthropic considerations are employed.
However, in terms of conformal time, $\eta$, the scale factor in 
the $\Lambda$-dominated phase is readily obtained by integration 
of Eq. \eqref{eq:2.2}, $a\propto(\eta_{*}-\eta)^{-1}$, 
where $\eta_{*}>0$ is a constant of integration. 
In that case $0<\eta<\eta_{*}$ is always finite, as is 
already evident from the discussion in section \ref{sec:2}.

From Eq. \eqref{eq:2.2} it follows 
that $d\eta=(\frac{8\pi G\rho}{3})^{-\frac{1}{2}}e^{-\xi}d\xi$, 
and so if we consider a flat prior in $\eta$ for observing 
the Universe, then the probability functions 
$\mathcal{P}(\eta)$ and $P(\xi|\{\Omega_{i}\})$ are 
related via
\begin{eqnarray}\label{eq:3.2}
\mathcal{P}(\eta)d\eta=C d\eta=P(\xi|\{\Omega_{i}\})d\xi
\propto\frac{d\xi}{\mathcal{H}(\xi;\{\Omega_{i}\})}
\propto\frac{e^{-\xi}d\xi}{\sqrt{\rho}},
\end{eqnarray}
where $\mathcal{H}$ is the conformal Hubble function, 
$\{\Omega_{i}\}$ stands collectively for the energy densities 
of the various contributions to the total energy budget 
(in critical density units), 
and $P(\xi|\{\Omega_{i}\})\Delta\xi$ is the 
conditional probability of making an observation 
over an interval 
$\Delta\xi$ around $\xi$ ($a\equiv e^{\xi}$) given 
the parameter values $\{\Omega_{i}\}$. 
Therefore, a flat prior on observing the Universe at around 
any given $\eta$, i.e. $\mathcal{P}(\eta)=C$, 
{\it generally} does not 
correspond to a flat prior on $\xi$, 
since $P(\xi|\{\Omega_{i}\})\propto\mathcal{H}^{-1}$  
varies with $\xi$ and is the largest (and 
correspondingly more likely for the 
Universe to be observed at) when the conformal expansion rate 
attains a minimum, i.e. $d\mathcal{H}/d\xi=0$, which corresponds 
by virtue of the Friedmann equation to $\mathcal{W}=-1/3$, where $\mathcal{W}$ is the density-weighted average EOS. 
In our case $\mathcal{W}=w\Omega_{DE}$, i.e. $\Omega_{DE}=-1/(3w)$.
Since $\mathcal{W}=-1/3$ is a mixture of both NR matter and DE 
then it follows that $w<-1/3$, in accordance with the discussion 
following Eq. \eqref{eq:2.5}. We reiterate that the EOS of the individual species 
are assumed to be fixed constants. Since $\mathcal{W}$ drops with 
the expansion of the Universe it follows from the currently observed  value, $\mathcal{W}=-2/3$, that we observe the Universe past its 
lowest conformal expansion rate, as is also discussed below 
Eq. \eqref{eq:2.6}.  

Applied to $w$CDM, Eq. \eqref{eq:3.2} reads
\begin{eqnarray}\label{eq:3.3}
P(\xi|\Omega_{m})d\xi &\propto &
\frac{d\xi}{\sqrt{\Omega_{m}e^{-\xi}
+(1-\Omega_{m})e^{-(1+3w)\xi}}}, 
\end{eqnarray}
where $w<-1/3$. It follows from Eq. \eqref{eq:3.3} that 
$P(\xi|\Omega_{m})$ peaks at $\xi=0$ ($a=1$) if $\Omega_{m}=(1+3w)/(3w)$
[which implies that if $w\lesssim -1/3$, as we have shown in section \ref{sec:2} to be necessary, then $\Omega_{m}\lesssim 1$, 
i.e. DE is never negligible at the most likely observable value].
The special case $\Omega_{m}=0$ and $w=-1/3$ corresponds to a flat 
distribution. This is the single case where both $\mathcal{P}(\eta)$ 
and $p(\xi)$ are flat, and the corresponding spacetime is that of Milne's empty spacetime \cite{116, 117, 118}. The lesson is that once $\rho\neq 0$ and the energy budget contains a DE component with $w<-1/3$ then $P(\xi|\Omega_{m})$ 
has a peak and is normalizable (first indication for the latter 
condition have been discussed in section \ref{sec:2} and will be 
further discussed below). In other words, $P(\xi|\Omega_{m})$ peaks 
solely due to the effect of gravitation, which is accounted for 
in the present context by invoking of the Friedmann equation 
in Eq. \eqref{eq:3.2}.

\begin{figure}[h]
\begin{center}
\leavevmode
\includegraphics[width=0.5\textwidth]{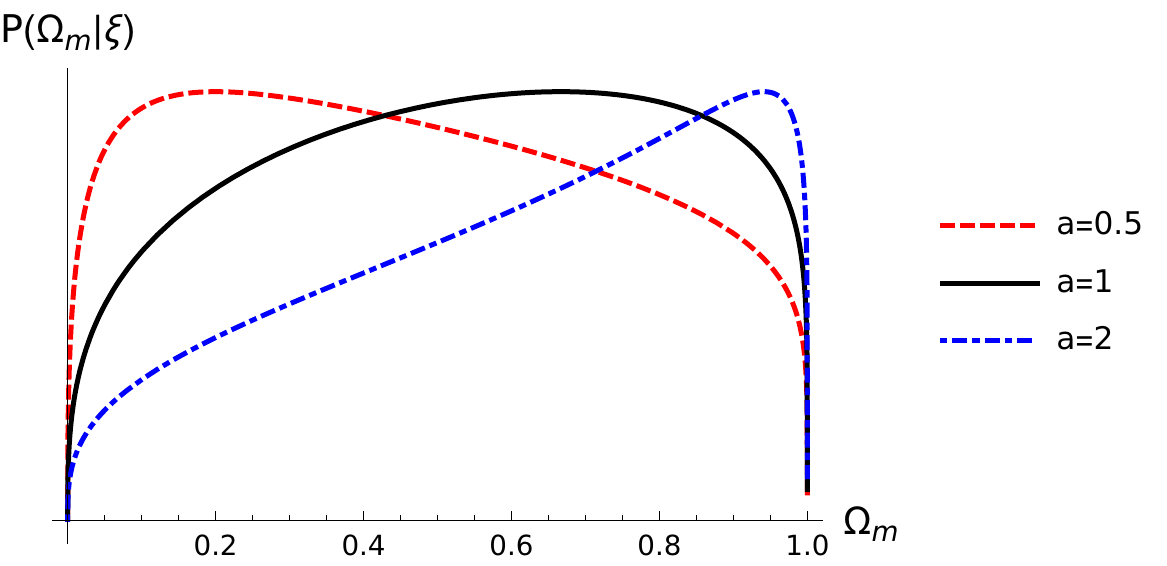}
\end{center}
\caption{\label{fig:3}
Shown is $P(\Omega_{m}|\xi)$, i.e. given that we observe 
at $a=0.5$, $1$ and $2$, what are the probability functions 
of $\Omega_{m}$ (dashed red, continuous black and dot-dashed blue curves, respectively).}
\end{figure}

Specifying to the case $w=-1$, 
i.e. assuming that DE is a pure cosmological constant, and 
viewing Eq. \eqref{eq:3.3} as a {\it normalized} 
distribution in $\xi$, we obtain
\begin{eqnarray}\label{eq:3.4}
\tilde{P}(\xi|\Omega_{m})=\frac{3\sqrt{\pi}\Omega_{m}^{\frac{1}{3}}
(1-\Omega_{m})^{\frac{1}{6}}}
{\Gamma(\frac{1}{3})\Gamma(\frac{1}{6})}
\frac{1}{\sqrt{\Omega_{m}e^{-\xi}+(1-\Omega_{m})e^{2\xi}}},
\end{eqnarray}
where $\int_{0}^{\infty}\tilde{P}(\xi|\Omega_{m})d\xi=1$. 
The steep decline of the probability function towards zero 
in the limits $\Omega_{m}\rightarrow 0$ and 
$\Omega_{m}\rightarrow 1$ 
that is evident from Eq. \eqref{eq:3.4} is visually 
illustrated in Fig.~\ref{fig:3}. 
The fact that when either 
$\Omega_{m}\rightarrow 0$ or $1$ the probability function 
vanishes basically implies that the fundamental question of why does 
DE exist at all (section \ref{sec:1}) is basically answered; 
under the assumption that we are typical along 
the $\eta$-axis, its maximum lifetime $\eta_{-1}$ has 
to be finite (section \ref{sec:2}), and observing the Universe is 
then only sensible if $\int_{0}^{\infty}\tilde{P}(\xi|\Omega_{m})d\xi=1$, 
which implies that DE is ought to exist. In fact, convergence 
of the integral only requires the existence of a phase of accelerated 
expansion, i.e. $w<-1/3$. Again, this is not independent 
of a similar conclusion arrived at in section \ref{sec:2}.

Interpreted in a Bayesian fashion, and applying Bayes Theorem to the case discussed here, results in a few useful 
results. The Bayes Theorem reads
\begin{eqnarray}\label{eq:3.5}
P(\xi|\Omega_{m})=\frac{P(\Omega_{m}|\xi)P(\xi)}{P(\Omega_{m})}.
\end{eqnarray}
Considering $\tilde{P}(\xi|\Omega_{m})
=P(\xi|\Omega_{m})P(\Omega_{m})$, the various 
probability functions can be readily read off. These are
\begin{eqnarray}\label{eq:3.6}
P(\Omega_{m}|\xi)&\propto&
\frac{\Omega_{m}^{\frac{1}{3}}
(1-\Omega_{m})^{\frac{1}{6}}}{\sqrt{\Omega_{m}e^{-\xi}
+(1-\Omega_{m})e^{2\xi}}},\\
\label{eq:3.7}
P(\Omega_{m})&=&\frac{\Omega_{m}^{\frac{1}{3}}
(1-\Omega_{m})^{\frac{1}{6}}}{B(\frac{4}{3},\frac{7}{6})},\\
\label{eq:3.8}
P(\xi)&=&\frac{1}{\xi_{2}-\xi_{1}},
\end{eqnarray}
and $P(\xi|\Omega_{m})$ follows from Eqs. \eqref{eq:3.4} 
and \eqref{eq:3.7}, and $B(\alpha,\beta)\equiv\Gamma(\alpha)\Gamma(\beta)/\Gamma(\alpha+\beta)$ is the Beta function of 
two parameters $\alpha$ and $\beta$.
The parameters $\xi_{2}$ and $\xi_{1}$ are upper and lower cutoffs on $\xi$ which are computationally irrelevant insofar they are sufficiently distant from $-\xi_{0}\equiv\frac{1}{3}\ln[2R]$ (i.e. $\xi_{1}\ll\xi_{0}\ll\xi_{2}$) where 
the distribution $P(\xi|\Omega_{m})$ peaks. 
Interestingly, it follows from 
Eq. \eqref{eq:3.7} that $\Omega_{m}$ is `sampled' from 
a `Beta distribution' and by virtue of the constraint 
$\Omega_{DE}=1-\Omega_{m}$ (ignoring curvature and radiation) it then follows that $\Omega_{DE}$ is drawn from an identical distribution. 
The flat prior on $\xi$ 
[Eq. \eqref{eq:3.8}] follows from applying the `principle of transformation groups', merely a generalization of the principle of indifference, to the scale factor $a$. The logarithm of $a$, i.e. $\xi$, 
becomes a `location parameter', for which the principle 
of transformation groups suggests a unique, uniform, prior.

Fig.~\ref{fig:3} delineates the probability distribution 
$P(\Omega_{m}|\xi)$ in the $\Lambda$CDM model 
for the cases $a=0.5$, $1$ and $2$, 
respectively. It is apparent from the plot that the distribution 
curves skew towards higher (lower) $\Omega_{m}$ values 
as $a$ is larger (smaller) than unity. This is easy to understand 
as $\Omega_{m}$ are the present day values, i.e. it is normalized to 
$a=1$. For example, if it is given that an observation is made at $a=2$ the observed value $\Omega_{m}(a=2)$ must match to the most probable value but since  
$\Omega_{m}(a)=\Omega_{m}a^{-3}/(\Omega_{m}a^{-3}+\Omega_{DE})$, 
i.e. $\Omega_{m}(a)$ decays in an expanding Universe, then the 
inferred $\Omega_{m}(a=1)\equiv\Omega_{m}$ is larger. 
Analogously, the observed $\Omega_{m}$ at present corresponds to larger values at any $a<1$.

Since $\Omega_{m}$ and $\Omega_{DE}$ appear with small powers 
in $P(\Omega_{m})$ the distribution function is broad, 
weakly dependent on, e.g. $\Omega_{m}$. To obtain the 
68, 95 and 99\% confidence intervals we calculate the area under the 
curve $P(\Omega_{m})$ leaving out the 16, 2.5 and 0.5\% 
distribution tails on each side of the distribution. In these 
cases we obtain
\begin{eqnarray}\label{eq:3.9}
\Omega_{m}&\in&[0.22,0.84]\ \ \&\ \ \Omega_{DE}\in[0.16,0.78];\ \ 68\% C.L.\nonumber\\
\Omega_{m}&\in&[0.055,0.968]\ \ \&\ \ \Omega_{DE}\in[0.032,0.945];\ \ 95\% C.L.\nonumber\\
\Omega_{m}&\in&[0.016,0.992]\ \ \&\ \ \Omega_{DE}\in[0.008,0.984];\ \ 99\% C.L.,
\end{eqnarray}
assuming no prior knowledge about cosmological parameters, 
other than that $\{\Omega_{i}\}$ are `sampled' from (0,1), 
and that we observe at the `present' time, i.e. at 
$\xi=0$. For reference, the concordance values, 
$\Omega_{m}=0.31$ and $\Omega_{DE}=0.69$, 
lie within the 68\% confidence interval.
 
\begin{figure}[h]
\begin{center}
\leavevmode
\includegraphics[width=0.5\textwidth]{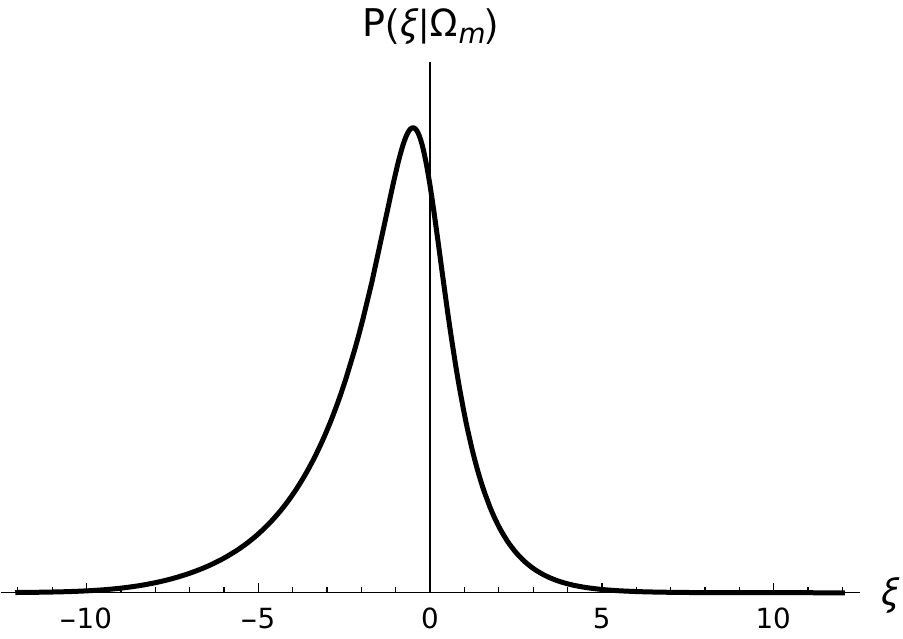}
\end{center}
\caption{\label{fig:4}
Likelihood for observation in the $\Lambda$CDM model 
with $\Omega_{DE}=0.69$ and $\Omega_{m}=0.31$. 
Within a few decades in $a$ we expect 
that at the time of observation $R=O(1)$.}
\end{figure}

\begin{figure}[h]
\begin{center}
\leavevmode
\includegraphics[width=0.5\textwidth]{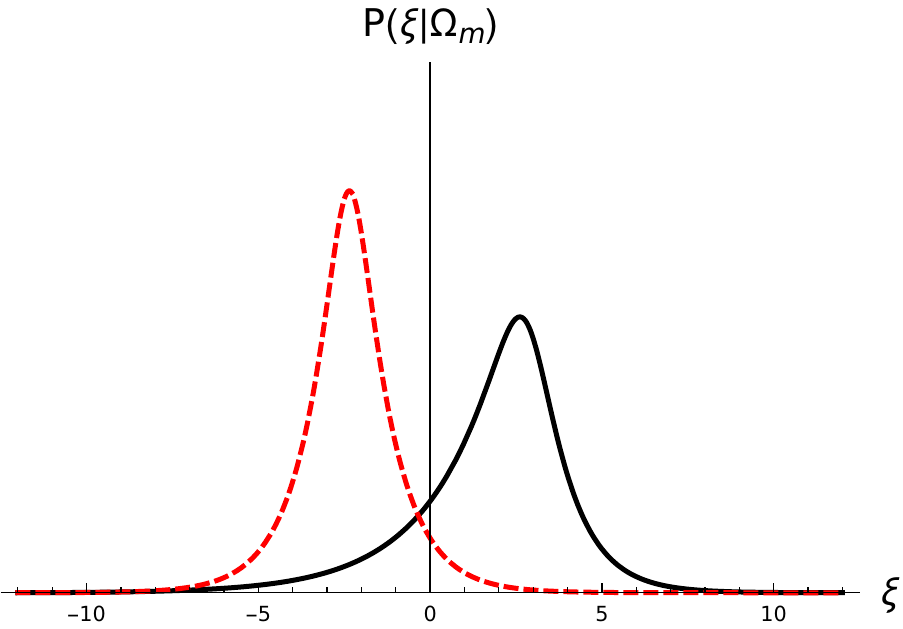}
\end{center}
\caption{\label{fig:5}
As in Fig.~\ref{fig:4} but with the $\Lambda$CDM 
parameters rescaled to the cases $a=1/20$ 
(black solid line) and $a=20$ (red dashed line). Comparison with 
Fig.~\ref{fig:4} illustrates that it is more likely to make a first 
observation of the 
Universe at the present time than it was at, e.g. 
$z=19$, or will be when the Universe grows twenty times larger than 
its current size.}
\end{figure}

To illustrate the robustness of our conclusion we 
show $P(\xi|\Omega_{m})$ at $\Omega_{m}=0.31$ in Fig.~\ref{fig:4}. Once again it is seen that 
the Universe is most likely to be observed 
when $R=O(1)$ within a relatively narrow range in $\xi$ 
[$P(\xi|\Omega_{m})$ has a clear peak and width while 
$\mathcal{P}(\eta)$ is constant and featureless]. 
Even though Fig.~\ref{fig:4} compellingly illustrates 
that the concordance $\Lambda$CDM is consistent with observations 
made at (or near) the peak of the probability function, it still remains to be checked for the {\it look-elsewhere effect}; how unique is 
the concordance $\Lambda$CDM in 
having the probability function peak 
at the present time? To test for the robustness of 
this result we consider 
the likelihood of observing the Universe at other values of $\xi$. Employing 
$\Omega_{m}(a)=\Omega_{m}a^{-3}/(\Omega_{m}a^{-3}+\Omega_{DE})$ 
and $\Omega_{DE}(a)=1-\Omega_{m}(a)$ where 
$\Omega_{m}\equiv\Omega_{m}(a=1)$, and 
$\Omega_{DE}\equiv\Omega_{DE}(a=1)$, 
and specifically considering the cases $a=1/(20)$ and $a=20$ results in Fig.~\ref{fig:5}.

To make contact with section \ref{sec:2}, and in particular to compare the present analysis to the constraints obtained in section \ref{sec:2} on 
$R$, we recast Eq. \eqref{eq:3.7}, using $\Omega_{m}=(1+R)^{-1}$, i.e. $P(\Omega_{m})d\Omega_{m}\propto P(R) dR$ where 
\begin{eqnarray}
P(R)=\frac{27\sqrt{\pi}}{2\Gamma(\frac{1}{3})\Gamma(\frac{1}{6})}R^{\frac{1}{6}}(1+R)^{-\frac{5}{2}}. 
\end{eqnarray}
This is a Beta Prime distribution (sometimes referred to as Beta distribution of the Second Kind or Inverted Beta distribution), with a probability distribution function of the form
$P(R)=R^{\alpha-1}(1+R)^{-\alpha-\beta}/B(\alpha,\beta)$, 
with $\alpha=7/6$ and $\beta=4/3$.

\begin{figure}[h]
\begin{center}
\leavevmode
\includegraphics[width=0.5\textwidth]{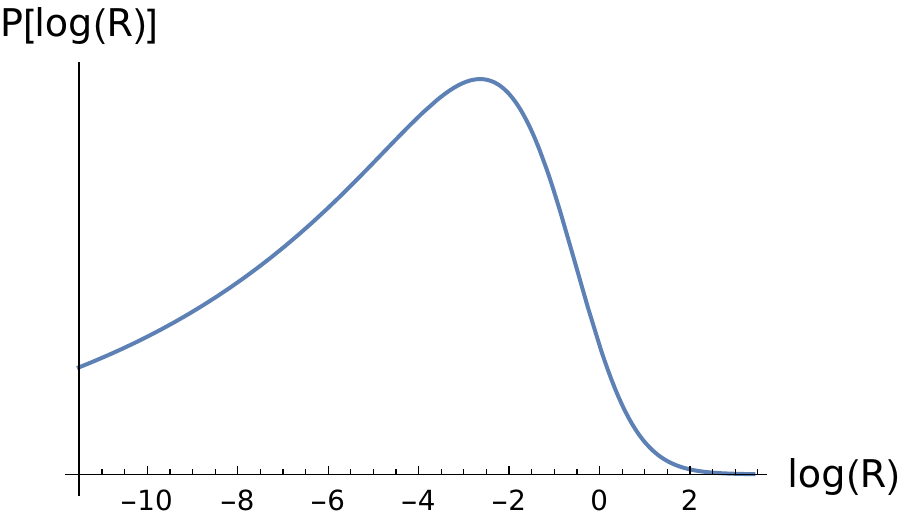}
\end{center}
\caption{\label{fig:6}
Shown is $P(R)$, the probability distribution function 
for $R=\Omega_{DE}/\Omega_{m}$, at the present time, 
on a semilogarithmic plot.}
\end{figure}

It is clear from its functional dependence that tight ({\it a priori}) 
upper limits on $R$ are placed once $R\gg 1$.
This probability distribution 
function for the dimensionless $R=\Omega_{DE}/\Omega_{m}$ 
results in the 50\%, 68\% and 95\% confidence intervals $0.32<R<2.17$, $0.20<R<3.46$ and $0.033<R<17.20$, respectively. The latter have been extracted from $P(R)$ 
following the prescription outlined above Eq. \eqref{eq:3.9}. 
The average value $\bar{R}=7/2$ is obtained 
from $\bar{R}=\int_{0}^{\infty}RP(R)dR$.
The probability distribution function $P(R)$ 
is shown in Fig.~\ref{fig:6} 

Again, in arriving at these confidence ranges no prior knowledge other than a uniform prior on $\eta$, in addition to the validity of the Friedmann equation 
(neglecting radiation and curvature and positing that $w=-1$), has been assumed. The observationally inferred value, $R_{obs}=2.23$, falls within the 68\% confidence range.

As mentioned below Eq. \eqref{eq:3.3}, 
it is demonstrated in Figs.~\ref{fig:4}-\ref{fig:5} 
that $P(\xi|\Omega_{m})$ is narrower than 
a flat $\mathcal{P}(\eta)$, and in addition the 
former peaks at around $\xi=0$ 
only in case that $\Omega_{m}$ is comparable to $\Omega_{DE}$; 
the CCP consequently arises from the (implied) expectation that 
we should {\it a priori} 
be positioned randomly along the {\it cosmic} timeline 
(impossible in a Universe with an infinite lifetime) 
while a more sensible choice would favor a uniform 
prior on $\eta$. In other words, our default expectation that we 
are `randomly sampled' from a uniform prior distribution on the 
expanding spacetime volume, rather than the comoving spacetime volume, inevitably results in the CCP, along with a few other hitherto puzzling (and even disturbing \cite{121, 115}) properties of the standard cosmological model.

\section{Other Implications}
\label{sec:4}
We point out two additional implications.  
First, so far we have considered the background 
evolution; we now discuss an implication for the growth of perturbations in the linear 
regime on sub-horizon scales (ignoring relativistic corrections). 
In the standard cosmological model, the matter overdensity, 
$\delta_{\rho}\equiv\delta\rho/\rho$, scales linearly with $a$, i.e. grows linearly 
with the scale factor inasmuch as the cosmic evolution is dominated by NR matter, e.g. \cite{101}. 
This steady growth stops once DE takes over. In our Universe, 
the latter transition took place fairly recently.
This well-understood growth dynamics, by itself, 
provided the basis for the proposed anthropic resolution 
of the CCP, e.g. \cite{2, 4}. However, 
as mentioned above already, rather than approaching 
the CCP via the growth rate of linear 
perturbations we attempt at employing 
the non-anthropic observational 
selection effect, that was described above, in explaining the 
current growth rate of linear density perturbations.

The overdensity, $\delta_{\rho}$, 
and scalar metric perturbations $\Phi$ in the gravitational 
potential, are related via the Bardeen equation and the 
relativistic generalization of the gravitational Poisson equation, e.g. \cite{119,120}
\begin{eqnarray}\label{eq:4.1}
\Phi''+3(1+\mathcal{W})\mathcal{H}\Phi'+k^{2}\Phi&=&0\nonumber\\
\mathcal{H}\Phi'+(\mathcal{H}^{2}+\frac{k^{2}}{3})\Phi=
-\frac{1}{2}\mathcal{H}^{2}\delta_{\rho},
\end{eqnarray}
where primes denote derivatives with respect to 
conformal time, $k$ is the Fourier mode of the perturbation, 
and $\mathcal{W}$ is the effective EOS of the {\it total} cosmic energy. 
Combining the Bardeen equation with the temporal derivative of the 
Poisson equation to eliminate the $\propto \Phi''$ term 
we obtain
\begin{eqnarray}\label{eq:4.2}
\mathcal{H}'(\Phi'+2\mathcal{H}\Phi+\mathcal{H}\delta_{\rho})
+k^{2}\Big[\frac{\Phi'}{3}-\mathcal{H}\Phi\Big]+
\frac{1}{2}\mathcal{H}^{2}\delta'_{\rho}=[2+3\mathcal{W}]\mathcal{H}^{2}\Phi'.
\end{eqnarray}
In the long wavelength limit, $k/\mathcal{H}\ll 1$, and at the most likely 
time for the Universe to be observed at, i.e. when $\mathcal{H}'=0$ 
(as concluded in section \ref{sec:3}), the first two terms 
on the left hand side drop 
and we are left with $\delta'_{\rho}=2(2+3\mathcal{W})\Phi'$. 
This implies that insofar $\mathcal{W}=-2/3$ the overdensity
$\delta_{\rho}$ attains a maximum at the most likely time for observation. 
For example, in case of the concordance cosmological model 
($\Omega_{m}\lesssim 1/3$ and $\Omega_{DE}\gtrsim 2/3$) 
$\mathcal{W}\approx -2/3$ 
we observe the Universe at its maximum 
overdensity on super-Hubble scales. 
Consequently, the most likely state of 
the Universe to be observed at 
is also the most clumpy one (at linear order). 
This sheds light on the anthropic approach to 
the CCP from a different perspective; the observed 
Universe is the one we see 
not necessarily because we require sufficient clustering for 
our existence, but rather because, statistically, 
the Universe is observed in its most likely 
state. This also turns out (but not {\it a priori} required) 
to be its clumpiest state 
(to first order and in the long wavelength limit), 
and is then likely to harbor sentient life.

The second implication is that 
from the perspective adopted in the present work, 
a spatially closed Universe (with $\Omega_{k}<0$) which contains 
only NR matter will most probably be observed at turnaround, when its expansion rate, 
$\mathcal{H}(\xi)=H_{0}\sqrt{\Omega_{m}e^{-\xi}-|\Omega_{k}|}$,
`momentarily' vanishes. 
A similar conclusion holds for a Universe 
that contains only matter and 
a {\it negative} cosmological constant, when its 
expansion rate, $\mathcal{H}(\xi)=H_{0}\sqrt{\Omega_{m}e^{-\xi}
-|\Omega_{\Lambda}|e^{2\xi}}$, vanishes.

\section{Discussion}
\label{sec:5}
The Copernican Principle applied to the standard cosmological model 
(i.e. the Cosmological Principle) proved to be a powerful simplifying assumption which to date has been found to be remarkably 
consistent with observations; on cosmological scales, 
there simply is no special point in space. 
However, when applied to the {\it cosmic} timeline, 
the Principle of Indifference -- the expectation that we are 
typical observers along the cosmic history -- 
results in a series of paradoxes and naturality problems.

The Cosmic Coincidence puzzle of the near equality of DE and NR matter contributions 
to the cosmic energy budget at present, arises from the 
prior expectation that, barring anthropic considerations, 
there seems to be no apparent 
reason (other than fine-tuned initial conditions) 
for their equality specifically at the time of observation, 
i.e. the present time, given their very different evolutionary histories. 

Irrespective of ones judgement of the scientific 
basis for the frequently-adopted assumption 
of typicality in cosmology, multiverse, planetary science, 
etc., the `why now?' problem associated with DE, 
by its very essence, relies on the implicit assumption of 
typicality; `why now' if no other time is {\it a priori} 
less favorable to witness the currently observed near 
equality of DE and NR? In other words, should the principle of typicality be discarded then the `why now?' problem automatically goes a way with it. 
In addition, as discussed in section \ref{sec:2} temporal typicality 
is merely a matter of definition (under certain plausible assumptions), 
at least in the cosmological context; What time coordinate are we typical at? 

However, as argued in the present work, 
accepting the cosmic coincidence as an {\it a priori} 
puzzling property of the standard cosmological model 
that needs to be explained, and assuming a flat prior 
for observing the Universe at any conformal time 
interval $\Delta\eta$ around $\eta$ seems to 
naturally explain the (otherwise) surprising 
comparability of DE and NR matter at present.
Specifically, a flat prior on $\eta$ does not correspond 
to a flat prior on $\xi$. It is 
an immediate consequence of the Friedmann equation 
that the probability distribution, $P(\xi|\Omega_{m})$, 
peaks at the minimum of the total energy density 
(as defined in the comoving frame). This corresponds 
to $R=O(1)$ and to the maximum (conformal) Hubble radius, 
$\mathcal{H}^{-1}$.

The essence of these considerations is that, in general, 
the total energy density $a^{2}\rho(a)$ 
(in the comoving frame) attains a minimum when 
the energy densities of DE and NR matter are comparable. 
Without having to introduce new fields or to change 
the dynamics of concordance $\Lambda$CDM, 
this provides a non-anthropic resolution to the CCP;
it implies that at a given randomly chosen finite interval 
$\Delta\xi$ the Universe is most likely 
to be found in a state where $R=O(1)$. 
All we need for that to work is to adopt the 
prior assumption that we are typical observers 
in the comoving spacetime volume.

As discussed in section \ref{sec:4}, the conclusion 
that the Universe is most likely to be observed 
at its lowest $a^{2}\rho(a)$, 
i.e. at its lowest (conformal) expansion rate, $\mathcal{H}$, 
elucidates the role played by anthropic 
considerations in proposed solutions of the problem. This is 
synonymous to overdensities at the long wavelength limit attaining 
their maximum value (to first order) when the background 
(conformal) density is minimized and the effective EOS 
is $\mathcal{W}=-2/3$, i.e. at an epoch very similar to 
the one we live in at the present era. 
The upshot is that a uniform prior for observing 
the Universe at any given time 
is in general consistent with the anthropic principle applied 
to this particular CCP. In other words, 
rather than imposing the requirement that the state of 
the Universe should be conducive to the existence of sentient beings 
to observe it, one can make the more modest (and perhaps natural) 
requirement that we are {\it a priori} equally likely to observe the Universe at any spacetime event in the four-dimensional 
{\it comoving} frame. 

\section*{Acknowledgements}
The author is indebted to Yoel Rephaeli for constructive and useful 
discussions. Comments by an anonymous referee 
on an earlier version of this work, that ultimately led to a 
significant expansion of this work, are greatly acknowledged.
This research has been supported by a grant 
from the Joan and Irwin Jacobs donor-advised fund at the 
JCF (San Diego, CA).

\end{document}